\documentclass[aps,amsmath,amssymb,amsfonts,twocolumn,showpacs]{revtex4-1}
\usepackage{bm,graphicx,mathrsfs}

\usepackage{graphicx}
\usepackage{epsfig}
\usepackage{amsmath,bbm}
\usepackage{amsfonts,amssymb}
\usepackage{times}
\usepackage{color}

\newcommand{\openPOVM}{\left(}
\newcommand{\closePOVM}{\right)}

\newcommand{\ii}{\mathbbm{1}}
\newcommand{\1}{\mathbbm{1}}

\newcommand{\tr}{{\rm tr}}

\newcommand{\be}{\begin{equation}}
\newcommand{\ee}{\end{equation}}
\newcommand{\bea}{\begin{eqnarray}}
\newcommand{\eea}{\end{eqnarray}}

\def\>{\rangle}
\def\<{\langle}
\def\|{\vert}
\newcommand{\hi}{\mathcal{H}}

\newcommand{\eh}{\mathcal{E}(\hi)}
\newcommand{\A}{{A}}
\newcommand{\B}{{B}}

\newcommand{\ran}{\textrm{ran}\,} 

\definecolor{james}{rgb}{1,.6,0}

\newcommand{\qed}{}

\def\Proof{\noindent{\it Proof: }}

\def\qed{\leavevmode\unskip\penalty9999 \hbox{}\nobreak\hfill
     \quad\hbox{\leavevmode  \hbox to.77778em{%
               \hfil\vrule   \vbox to.675em%
               {\hrule width.6em\vfil\hrule}\vrule\hfil}}
     \par\vskip3pt}

\newtheorem{theorem}{Theorem}

\newtheorem{proposition}{Proposition}
\newtheorem{definition}[theorem]{Definition}

\begin{document}

\author{David Reeb}
\email{david.reeb@tum.de}
\author{Daniel Reitzner}
\email{danreitzner@gmail.com}
\author{Michael M.\ Wolf}
\email{m.wolf@tum.de}

\affiliation{Department of Mathematics, Technische Universit\"at M\"unchen, 85748 Garching, Germany
}

\title{Coexistence does not imply joint measurability}


\begin{abstract}
One of the hallmarks of quantum theory is the realization that distinct measurements cannot in general be performed simultaneously, in stark contrast to classical physics. In this context the notions of \emph{coexistence} and \emph{joint measurability} are employed to analyze the possibility of measuring together two general quantum observables, characterizing different degrees of compatibility between measurements. It is known that two jointly measurable observables are always coexistent, and that the converse holds for various classes of observables, including the case of observables with two outcomes. Here we resolve, in the negative, the open question whether this equivalence holds in general. Our resolution strengthens the notions of coexistence and joint measurability by showing that both are robust against small imperfections in the measurement setups.
\end{abstract}

\maketitle

It is well known that two quantum observables can in general not be measured together \cite{heisenbergur}. In describing the relation between two or more quantum observables, several related notions are in use. The most prominent ones are:\ commutativity ({\textsf{COM}}), non-disturbance ({\textsf{ND}}), joint measurability ({\textsf{JM}}), and coexistence ({\textsf{COEX}}) \cite{HeinosaariWolf,LudwigBook,Pulmannova1,Lahti,book}. Whereas, as the names suggest, joint measurability and non-disturbance can easily be understood in operational terms, commutativity and coexistence at first glance rely more on the underlying mathematical representation of quantum observables.

The connections between all these properties are well studied for pairs of general quantum observables, which are given in terms of \emph{positive operator-valued measures (POVM)}. If the POVMs are projection-valued --- the case considered in most undergraduate quantum physics textbooks --- then all four notions turn out to coincide, which may explain why they are sometimes used interchangeably. In general,  we know that 
$${\textsf{COM}}~\Rightarrow~{\textsf{ND}}~\Rightarrow~{\textsf{JM}}~\Rightarrow~{\textsf{COEX}}$$
holds, and that the first two implications are strict in the sense that the reverse implications do not hold in general \cite{HeinosaariWolf}.  The last implication, however, appears to be more subtle:\ while joint measurability is known to imply coexistence, it is a persistent open problem whether the converse holds as well \cite{Lahti,Jukka,Carmeli}. The present paper resolves this problem.

We begin with recalling the basic definitions and setting the notation. On a complex Hilbert space $\hi$, a linear operator $E$ with $0\leq E\leq\1$ is called an \emph{effect}. The set of effects is denoted by $\eh$. A general quantum \emph{observable} (or \emph{measurement}) is described by a POVM $A$, which is a countably additive mapping $\A:\mathcal A\to\eh$ on a $\sigma$-algebra $\mathcal A$ of subsets of $\Omega_A$ satisfying $\A(\Omega_A)=\1$. The set $\Omega_A$ represents the possible outcomes of the measurement. For any input state $\rho$ describing the initial preparation of the quantum system and for any $X\in\cal A$, the expression $\tr{\big[\rho A(X)\big]}$ gives then the probability of obtaining a measurement outcome $x\in X$ \cite{holevobook}. We denote by $\ran(\A):=\{\A(X)|X\in\mathcal A\}$ the set of effects corresponding to $A$.

For the results below it will be sufficient to consider finite outcome sets $\Omega_A=\{1,\ldots,n\}$ equipped with the discrete $\sigma$-algebra ${\mathcal A}$ that contains all subsets of $\Omega_A$; we call $A$ an \emph{$n$-outcome observable}. In this case, $A$ is fully determined by the effects $A_k:=A(\{k\})$ for $k\in\{1,\ldots,n\}$, and abusing notation we then write $A=\openPOVM A_1,\ldots,A_n\closePOVM$.

We now define the two notions whose relationship we want to clarify.
\begin{definition}[Coexistence]Two POVMs $\A$ and $\B$ are called \emph{coexistent} if there exists a POVM $M$ such that $\ran(\A)\cup\ran(\B)\subseteq\ran(M)$.\end{definition}
The notion of \emph{coexistence} was introduced for effects and for observables by Ludwig \cite{LudwigBook} and refined to the present definitions by Busch, Lahti and Mittelstaedt \cite{buschlahtimittelstaedt}. Coexistence of the two observables $A$ and $B$ ensures that each effect of $A$ or $B$ can be simulated by the measurement $M$, and even that all binary observables that can be formed from $A$ and $B$ can be measured simultaneously. But it does not directly provide a way to measure the entire observables $A$ and $B$ simultaneously. 

A simultaneous measurement is possible when $A$ and $B$ are both marginals of a single observable. This is captured by the following notion:\begin{definition}[Joint measurability]\label{defineJM}Two POVMs $\A:\mathcal A\to\eh$ and $\B:\mathcal B\to\eh$ are \emph{jointly measurable} if there exists a POVM $J:{\mathcal J}\to\eh$ on the $\sigma$-algebra ${\mathcal J}$ generated by ${\mathcal A}\times{\mathcal B}$, such that for all $X\in\mathcal A$ and $Y\in\mathcal B$,
\[
\A(X)\,=\,J(X\times\Omega_\B)~,\qquad \B(Y)\,=\,J(\Omega_\A\times Y)~.
\]
\end{definition}

Joint measurability of two observables immediately implies their coexistence. The converse also follows easily for binary observables: the two-outcome POVMs $\A=(A_1,\1-A_1)$ and $\B=(B_1,\1-B_1)$ are coexistent if and only if they are jointly measurable \cite{Pulmannova2,KrausBook,Lahti}.

Beyond this case of two outcomes, several broad classes of observables have been identified for which coexistence and joint measurability are equivalent \cite{Lahti,Pulmannova1,Pulmannova2,dvurecenskij,juhapekkapellonpaa}:\ for example projection-valued POVMs \cite{Pulmannova1}; all cases in which one of the POVMs is determined by a discrete set of linearly independent rank-1 effects, as noticed very recently \cite{juhapekkapellonpaa,footnotejuhapekka}; or POVMs with effects contained in a regular effect algebra \cite{Pulmannova2}. So far, however, it was an open question whether the equivalence holds for all pairs of observables \cite{Lahti,Jukka,Carmeli}.

\begin{figure}
\includegraphics[scale=0.9]{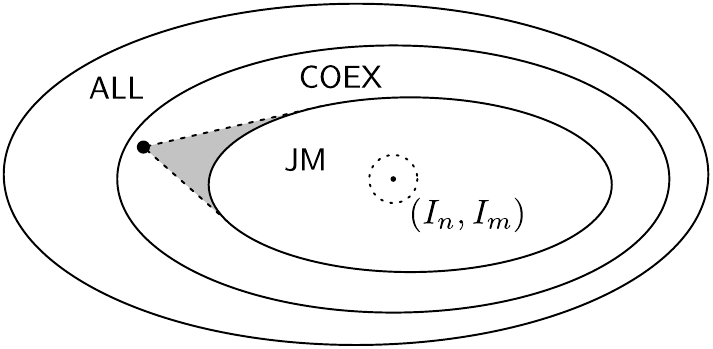}
\caption{\label{convexfig}An illustration of the compact convex sets of all pairs $(A,B)$ of $n$- resp.\ $m$-outcome observables ({\textsf{ALL}}), of coexistent pairs ({\textsf{COEX}}), and of jointly measurable pairs ({\textsf{JM}}). The set {\textsf{JM}} contains an open neighborhood around the pair $(I_n,I_m)$ of uniformly random observables and has thus positive volume. For $n\geq3$, $m\geq2$, the existence of the pair (solid dot) from Proposition \ref{counterexampleprop} implies that the set difference ${\textsf{COEX}}\!\setminus\!{\textsf{JM}}$ has positive volume (see shaded area), whereas ${\textsf{COEX}}={\textsf{JM}}$ whenever $n,m\leq2$ \cite{Pulmannova2,KrausBook,Lahti}. By similar reasoning, ${\textsf{ALL}}\!\setminus\!{\textsf{COEX}}$ has positive volume iff $n,m\geq2$.}
\end{figure}

We answer this question by providing an instance of coexistent observables that are not jointly measurable. The example has $|\Omega_A|=3$ and $|\Omega_B|=2$ and is thus minimal in terms of the number of outcomes beyond the known two-outcome case.

Let $\{|1\>,|2\>,|3\>\}$ be an orthonormal basis in $\hi=\mathbb{C}^3$ and $|\psi\>:=(|1\>+|2\>+|3\>)/\sqrt{3}$. Consider the following effects:
\begin{align*}
A_i\,&:=\,\frac12\big(\1-|i\>\<i|\big)\,,\quad i\in\{1,2,3\}\,,\\
B_1\,&:=\,\frac12\;|\psi\>\<\psi\|\,,\quad B_2\,:=\,\1-B_1\,.
\end{align*}
\begin{proposition}\label{counterexampleprop}
The POVMs $A:=\openPOVM A_1,A_2,A_3\closePOVM$ and $B:=\openPOVM B_1,B_2\closePOVM$ are coexistent, but not jointly measurable.
\end{proposition}  
\Proof To prove coexistence of $A$ and $B$, each of which has at most three outcomes, we have to construct a POVM whose range contains each $A_i$ and $B_j$. The 5-outcome observable
\begin{equation*}
M\,:=\,\openPOVM\frac12\|1\>\<1\|,\,\frac12\|2\>\<2\|,\,\frac12\|3\>\<3\|,\,B_1,\,\,\frac12\1-B_1\closePOVM
\end{equation*}
clearly does the job.
Concerning joint measurability we argue by contradiction. Suppose the observables $A$ and $B$ were jointly measurable. Then, by Definition \ref{defineJM}, there exist effects $J_{ij}\geq 0$ such that 
\bea \forall i:\ \ \sum_{j=1}^2 J_{ij}=A_i\,\text{,~~~~~and~~~}\forall j:\ \ \sum_{i=1}^3 J_{ij}=B_j\label{eq:JBj}\,.\eea
Since by Eq.\ (\ref{eq:JBj}) the positive-semidefinite operators $J_{i1}$ sum to the rank-1 operator $B_1$, we must necessarily have $J_{i1}=c_iB_1$ for some numbers $c_i\geq0$. Hence, again by Eq.~(\ref{eq:JBj}), 
$A_i=c_iB_1+J_{i2}$, which, after taking the overlap $\<i|\cdot|i\>$, becomes
\begin{equation*}
0\,=\,\frac{c_i}{2}\,|\<i|\psi\>|^2+\<i\|J_{i2}|i\>\qquad\forall i\in\{1,2,3\}~.
\end{equation*}
This implies $c_i=0$ for all $i$ due to $|\<i|\psi\>|^2=1/3$ and $\<i\|J_{i2}|i\>\geq0$. Then, however, $J_{i1}=0$ for all $i$, and Eq.\ (\ref{eq:JBj}) leads to the desired contradiction $B_1=\sum_i J_{i1}=0$.\qed

By padding both POVMs from Proposition \ref{counterexampleprop} with effects $0\in\eh$, one sees that for every $n\geq3$, $m\geq2$ there exist $n$- resp.\ $m$-outcome POVMs $A$ and $B$ that are coexistent but not jointly measurable.

Proposition \ref{counterexampleprop} enables a geometric picture of joint measurability and coexistence for pairs of observables on a fixed Hilbert space $\hi$ of finite dimension at least $3$. First consider the pair $(I_n,I_m)$, where $I_k:=(\ii/k,\dots,\ii/k)$ denotes the $k$-outcome POVM corresponding to the toss of an unbiased $k$-sided coin. Obviously, $I_n$ and $I_m$ are jointly measurable. Since $n$- resp.\ $m$-outcome observables $A$ and $B$ are jointly measurable whenever all their effects satisfy $A_i\geq\ii/2n$ and $B_j\geq\ii/2m$ \cite{schultzheinosaari}, any pair $(A,B)$ sufficiently close to $(I_n,I_m)$ is jointly measurable as well. Within the set {\textsf{ALL}} of all pairs of POVMs, the set {\textsf{JM}} of jointly measurable pairs has thus non-empty open interior, see Fig.\ \ref{convexfig}. By Definition \ref{defineJM}, {\textsf{JM}} is furthermore convex and closed in the direct sum space of all pairs. Closedness of {\textsf{JM}} is ensured for pairs of finite-outcome observables on the finite-dimensional $\hi$ since the joint observables $J$ from Definition \ref{defineJM} live then in a compact set.

The set {\textsf{COEX}} of coexistent pairs of $n$- resp.\ $m$-outcome POVMs is convex as well since coexistence of $A$ and $B$ is equivalent to the joint measurability of the collection of all binary POVMs that can be formed from $A$ and $B$. The preceding observations lead thus to the conclusion depicted in Fig.\ \ref{convexfig}: Since Proposition \ref{counterexampleprop} guarantees the existence of a pair $(A,B)\in{\textsf{COEX}}\!\setminus\!{\textsf{JM}}$ for $n\geq3$, $m\geq2$, the intersection of ${\textsf{COEX}}\!\setminus\!{\textsf{JM}}$ with the convex set spanned by $(A,B)$ and ${\textsf{JM}}$ has non-empty open interior. Therefore, the set ${\textsf{COEX}}\!\setminus\!{\textsf{JM}}$ itself has non-empty open interior, i.e.\ positive volume.

We emphasize that the latter conclusion resolves the question answered by this article in a strong sense: Whereas it was previously unknown whether there exists even one coexistent but not jointly measurable pair of observables, our Proposition \ref{counterexampleprop} implies that both notions are different and that this difference is not merely an exceptional or spurious effect. Rather, a positive fraction of all pairs of observables are jointly measurable, and another positive fraction are coexistent but not jointly measurable. This ensures stability features against small perturbations in the distinction between coexistence and joint measurability, which makes both notions more meaningful and robust in experimental setups.

\medskip


{\it Acknowledgments:} The authors thank Toby Cubitt, Teiko Heinosaari, and Jukka Kiukas for valuable comments.

\end{document}